# Electronic Transport Evidence for Topological Nodal-Line Semimetals of ZrGeSe single crystals

Lei Guo,[†,‡,∇] Ting-Wei Chen,[‡,∇] Chen Chen,[†] Lei Chen,[§,#] Yang Zhang,[†] Guan-Yin Gao,[⊥] Jie Yang,[&] Xiao-Guang Li,[⊥] Wei-Yao Zhao,[§,#,*] Shuai Dong,[†,*] and Ren-Kui Zheng[‡,§*]

[†] School of physics, Southeast University, Nanjing 211189, China

[‡] School of Materials Science and Engineering and Jiangxi Engineering Laboratory for Advanced Functional Thin Films, Nanchang University, Nanchang 330031, China

[§] State Key Laboratory of High Performance Ceramics and Superfine Microstructure, Shanghai Institute of Ceramics, Chinese Academy of Sciences, Shanghai 200050, China

[#] ISEM, Innovation Campus, University of Wollongong, Wollongong, NSW 2500, Australia

[⊥] Hefei National Laboratory for Physical Sciences at the Microscale, Department of Physics, and Collaborative Innovation Center of Advanced Microstructures, University of Science and Technology of China, Hefei 230026, China

[&] Suzhou Institute of Nano-Tech and Nano-Bionics, Chinese Academy of Sciences, Suzhou 215123, Jiangsu, China

**Abstract:** Although the band topology of ZrGeSe has been studied via magnetic torque technique, the electronic transport behaviors related to the relativistic Fermions in ZrGeSe are still unknown. Here, we first report systematic electronic transport properties of high-quality ZrGeSe single crystals under magnetic fields up to 14 T. Resistivity plateaus of temperature dependent resistivity curves both in the presence and absence of magnetic fields as well as large, non-saturating magnetoresistance in low-temperature region were observed. By analyzing the temperature- and angular-dependent Shubnikov-de Haas oscillations and fitting it via the Lifshitz-Kosevich (LK) formula with the Berry phase being taken into account, we proved that Dirac fermions dominate the electronic transport behaviors of ZrGeSe and the presence of non-trivial Berry phase. First principles calculations demonstrate that ZrGeSe possesses Dirac bands and normal bands near Fermi surface, resulting in the observed magnetotransport phenomena. These results demonstrate that ZrGeSe is a topological nodal-line semimetal, which provides a fundamentally important platform to study the

---

quantum physics of topological semimetals.

## 1. INTRODUCTION

Topological state of quantum materials is now one of the spotlights in condensed matter physics and materials science and has attracted tremendous attention as of now. Different types of band topology in quantum materials have been predicted and investigated since the first report of topological insulators a decade ago.[1,2] Under ideal conditions, the Fermi surface, which determines the electronic transport properties of topological quantum materials, is composed of isolated band crossing points and/or lines, which is different from those of normal conductors.[3,4] Because of the peculiar Fermi surface, topological quantum materials possess a plenty of quantum phenomena such as extremely large magnetoresistance, ultrahigh carrier mobility, quantized Hall effect, and chiral anomaly in magnetoresistance, which are not only fundamentally important but also provide potential device applications.[5,6] Besides bulk gapped topological insulators, bulk gapless topological semimetals have been progressively discovered in recent years. According to the degeneracy and distribution of band-crossing points in the Brillouin zone, topological semimetals are classified into Weyl semimetals, Dirac semimetals, and nodal-line semimetals.[7-11] Among them, Dirac semimetals usually regarded as a three-dimensional graphene, which possess band crossing points with four-fold degeneracy near the Fermi energy ($E_F$). Alternatively, the Dirac points degenerate to pairs of discrete Weyl points with double degeneracy and different chirality (relevant materials are denoted as Weyl semimetals).[7-9] For nodal-line semimetals, the band crossing forms closed loops instead of discrete points in Brillouin zone.[10-12] Theoretical calculations predicted that the spin-orbital coupling may lead the nodal ring evolves to Weyl nodes, Dirac nodes, or open a gap to form a topological insulator, thus making nodal-line semimetal become one of the fundamental materials in topological physics studies.[12,13]

Several single crystals have been predicted to be nodal-line semimetals, including $Ca_3P$,[14] $Ca_3P_2$,[15] $A_2B$ ($A$=Ca, Sr and Ba; $B$=As, Sb and Bi),[16] $Mg_3Bi_2$,[17] LaN,[18] $Cu_3$(Pd,Zn)N,[19,20] (Tl,Pb)TaSe$_2$,[21-22] ZrSiS,[23-27] CaAg(P,As),[28] Ba$MX_3$ ($M$=V, Nb, Ta; $X$=S, Se).[29] Unfortunately, experimental progresses on nodal-line semimetals are quite limited. Particularly, electronic transport evidences are only presented in very limited materials and directly confirmation of the existence of nodal rings are verified only in PbTaSe$_2$[22] and ZrSiS[27] compounds by angle-resolved photoemission spectroscopy (ARPES). Among the aforementioned semimetals, ZrSiS attracts the most attention for its distinguished band structure and interesting transport properties. In ZrSiS, all energy bands crossing the Fermi surface are Dirac bands (linearly dispersed over a larger energy range than any other quantum materials), which implies that one can study behaviors of pure Dirac electron without interference from normal electrons.[27] Further, ZrSiS harbors two types of unusual Dirac cones: the first one forms the known closed loop near the Fermi level and the other, generated by a square Si lattice protected by non-symmorphic symmetry, won't open a gap regardless of the strength of spin-orbit coupling.[27] Since ZrSiS is quite popular recently, it is necessary to investigate on some isostructural materials. Hu et al.[30] have studied the Fermi surface of the ZrSiSe and ZrSiTe via the de Haas-van Alphen (dHvA) quantum oscillations and deduced that both materials are nodal-line semimetal candidates. Similar method has been applied to study ZrGe$M$ ($M$=S, Se and Te)[31] and ZrSnTe[32] single crystals whose dHvA quantum oscillations witnesses the possible existence of nodal-line Fermions in this family of materials.

Although the dHvA quantum oscillations is an important technique for extracting the relativistic nature of Dirac fermions in topological quantum materials and has its own advantages (e.g., no carrier's scattering is involved). It is also important to understand the electronic transport properties of topological semimetals for the purpose of basic research and device applications which usually take advantage of electronic properties. Motivated by this, we significantly optimized the quality of ZrGeSe single crystals and systematically investigated the electronic transport properties of these single crystals which show resistance plateaus both in the absence ($H$=0 T) and presence of magnetic field (0<$H$≤14 T) and large non-saturating magnetoresistance in low-temperature region (3 K ≤ $T$ ≤ 50 K). By analyzing the Shubnikov-de Haas (SdH) oscillations of the magnetoresistance obtained at fixed low temperatures, we deduced that the nodal-line Fermions dominate the electronic transport behaviors and the presence of non-trivial Berry phase in ZrGeSe single crystals. On the other hand, the findings of large non-saturating magnetoresistance and high mobility in this Van de Waals single crystal may encourage more basic property studies and device fabrication researches.

## 2. EXPERIMENTAL SECTION

High-quality ZrGeSe single crystals were grown by the chemical vapor transport (CVT) method using the iodine ($I_2$) as transport agent.[30,31] Briefly, high-purity stoichiometric amount (~1.5 g) of Zr, Ge and Se powder (200 mesh), together with 10 mg/ml iodine, are sealed in quartz tube as the starting materials. The crystal growth was carried out in a two-zone furnace between 950 °C (source) and 850 °C (sink) for 1 week. The as-grown single crystals exhibit plate-like shape with a typical size of approximately 5 to 10 millimeter in length and 3 to 4 millimeter in width and the normal vector of [001].

The crystal structure of the single crystals were characterized by a PANalytical X'pert x-ray diffractometer equipped with Cu $K\alpha_1$ radiation. High-resolution transmission electron microscopy (HRTEM) and selected area electron diffraction (SAED) were measured using a Tecnai G2F20 S-Twin transmission electron microscope. Note that the electron beam is incident along the [110] crystallographic direction. The chemical composition and distribution maps of the single crystals were measured via energy dispersive x-ray spectroscopy (EDS) using an x-ray energy dispersive spectrometer (Oxford Aztec X-Max80) installed on the Zeiss Supra 55 scanning electron microscope.

The electronic transport properties were measured by the standard four-probe method using a physical property measurement system (PPMS-14T, Quantum Design). Ohmic contacts were prepared on a fresh cleavage *ab* plane using room-temperature cured silver paste. The electric current is parallel to the *b* axis while the direction of the magnetic field is parallel to the *c* axis of the crystal in the transverse magnetotransport measurement configuration.

The first-principles electronic structure calculations are performed using the Vienna *ab initio* simulation package (VASP) with the projector augmented-wave (PAW) potentials.[33,34] To acquire an accurate description of the crystalline structure, the Perdew-Burke-Ernzerhof for solids (PBEsol) functional is adopted.[35] The cutoff of the plane wave basis was fixed to 500 eV and the Monkhorst-Pack k-point mesh is set to be 11×11×5 for the minimal cell. Both the lattice constants and atomic positions are fully relaxed until the force on each atom is below 0.01 eV/Å. The Fermi surface was handled with Vaspkit, a post-VASP code, using a 54×54×24 k-mesh.[36]

## 3. RESULTS AND DISCUSSIONS

1）**Characterization of ZrGeSe Single Crystals**

Similar to other *WHM* compounds (*W*=Zr/Hf/La, *H*=Si/Ge/Sn/Sb, *M*=O/S/Se/Te), ZrGeSe possesses a layered tetragonal crystal structure formed by stacking of Se-Zr-Ge-Zr-Se slabs along the *c*-axis [**Fig. 1(a)**].[37] As a result of the layered crystal structure, the obtained ZrGeSe single crystals via the CVT method exhibit plate-like shape. A large and smooth *ab* plane and a layered structure near the edge of a crystal can be observed from the SEM images shown in **Fig. 1(b)** and **1(c)**, respectively. **Fig. 1(d)** shows a photograph of a typical ZrGeSe single crystal with centimeter in-plane size. The excellent crystalline quality of the single crystal is reflected by the sharp and clean comb-like (00*l*) (*l*=1, 2, 3, 4, 5, 6) X-ray diffraction (XRD) pattern [**Fig. 1(e)**] and a very small full width at half maximum (FWHM) (~0.04º) of the XRD rocking curve taken on the (003) diffraction peak [**Fig. 1(f)**]. High-resolution transmission electron microscopy (HRTEM) demonstrates that the microstructure of the single crystal is nearly perfect **[Fig. 1(g)]**, resulting in sharp selected area electron diffraction pattern **[Fig. 1(h)]** which can be indexed with the P4/nmm space group. X-ray energy dispersive spectroscopy (EDS) reveals uniform distribution of the Zr, Ge, and Se element and the atomic ratios of Zr:Ge:Se are 33.8 : 33.5 : 32.7 (**Fig. S1, Supporting Information**). All of these results demonstrate the excellent quality of the ZrGeSe single crystal which provides good platform to study its electronic transport properties.

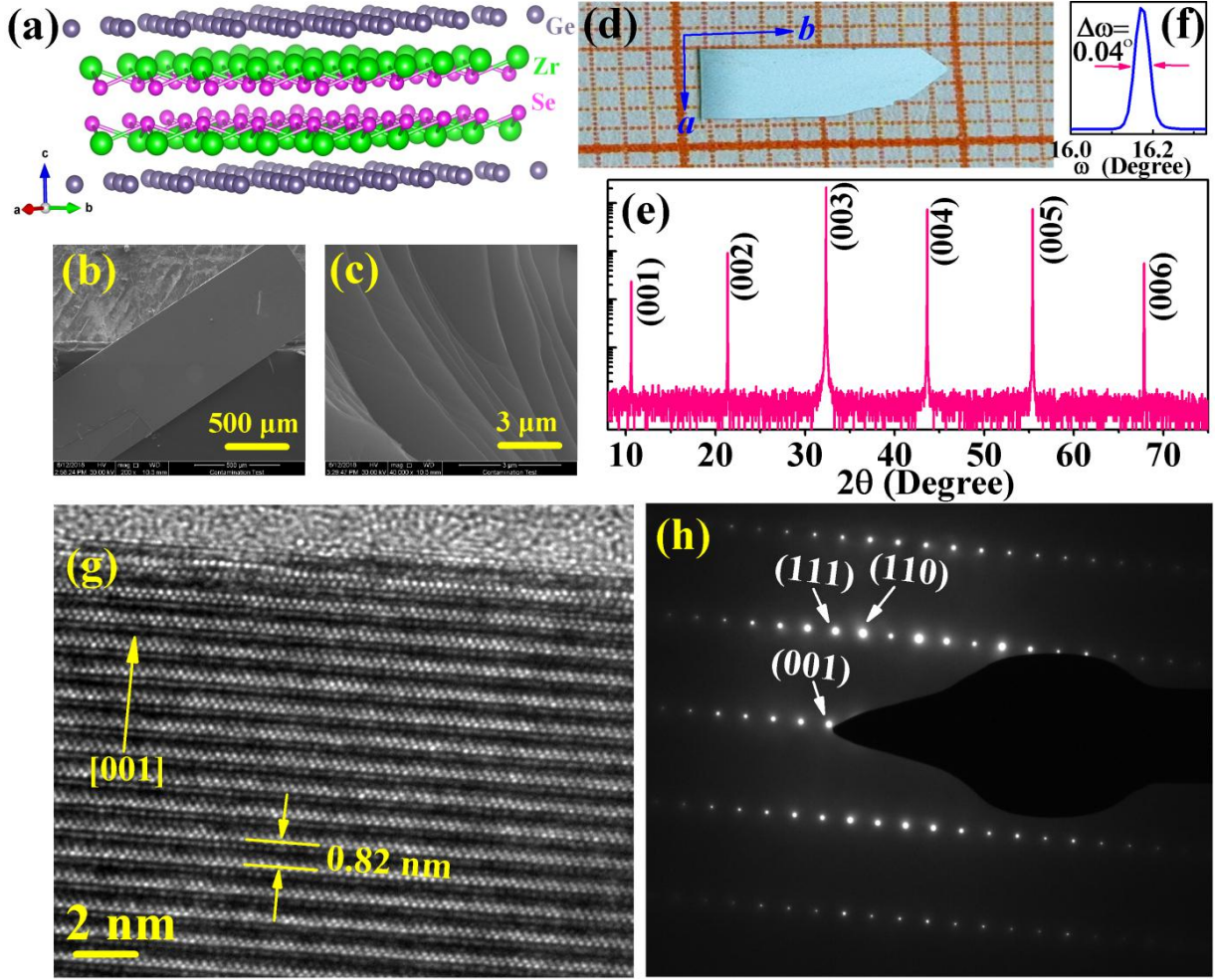

**Fig. 1 Structure characterization of ZrGeSe single crystals.** (a) Crystal structure of ZrGeSe. (b) SEM image of a ZrGeSe single crystal. (c) SEM image of a ZrGeSe single crystal taken near its edge. (d) A photograph of a typical ZrGeSe single crystal with 1 cm in length. (e) XRD patterns of the ZrGeSe. (f) XRD rocking curve taken on the ZrGeSe (003) peak. (g) A high-resolution TEM image. (h) Selected area electron diffraction pattern of the ZrGeSe single crystal.

As shown in **Fig. 2a**, the zero-field resistivity-temperature (*R-T*) curve of the ZrGeSe crystal shows monotonically increasing within the entire temperature region (3-300 K), which suggests the metallic ground state of the ZrGeSe. For $T \leq 10$ K the resistivity shows slight fluctuation, which is similar to that reported in $Cd_3As_2$[38] and ZrSiS[39] Dirac semimetals and could be explained in terms of the quantum ballistic transport. The residual resistivity ratio (RRR) is approximately 14.9 under a zero magnetic field, suggesting that ZrGeSe crystal possesses a good metallic property. Moreover, the resistivity in the low-temperature region ($T \leq 75$ K) can be quite well fitted using the equation $\rho(T)=\rho_0 +AT^n$ with $n \sim 3.1$ (**Fig. S2, Supporting Information**), where $\rho_0$ is the residual resistivity, $A$ is a constant, and $n$ is the parameter indicating scattering mechanisms. The obtained $n$ value (~3.1) is close to that of the ZrSiS crystal ($n \sim 3$).[39] This type of temperature dependence of resistivity, deviating from the pure electronic correlation-dominated scattering mechanism ($n=2$) observed in semimetal families,[40] can be attributed to the interband electron-phonon scattering.[39] In the presence of magnetic fields up to 14 T along the $c$ axis, the resistivity plateaus still remain for $T \leq 10$ K, which can be clearly seen by using the logarithmic temperature axis (**Fig. 2a**). Moreover, one can find that the application of magnetic fields induces a metal-to-insulator-like transition in the low-temperature region. Specifically, for $B \geq 3$T the resistivity shows a minimum at a certain temperature ($T_m$) below which the resistivity increases with decreasing temperature.

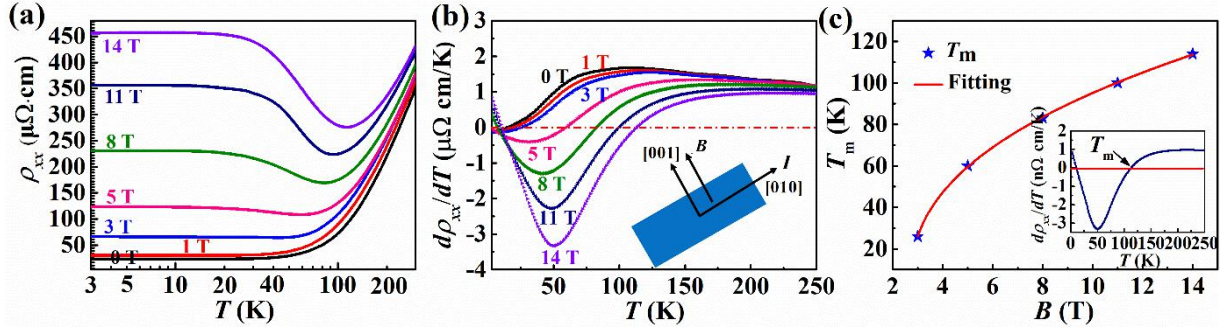

**Fig. 2. Temperature dependent transport properties of ZrGeSe single crystal.** (a) Temperature dependence of resistivity in different magnetic fields, as measured using the schematic geometry shown in the inset of (b). (b) $d\rho_{xx}/dT$ plotted as a function of temperature. (c) $T_m$ plotted as a function of magnetic field. Inset: $T_m$ value obtained from the differential curve.

We employ d$\rho_{xx}$/d$T$–$T$ plot to obtain the $T_m$ values which are the $T$ values at d$\rho_{xx}$/d$T$ = 0 in **Fig. 2b**. The obtained $T_{min}$ value increases monotonically with increasing magnetic fields, as shown in **Fig. 2c**, and can be fitted by the equation $T_{min} \sim (B-B_0)^{1/\nu}$. The parameter ν is ~2.7, which is close to that of ZrSiS ( ν =3) crystal.[39] This type of magnetic-field-induced metal-to-insulator-like resistivity transition sometimes may result from the gap opening at the band-touching points in semimetals.[39,41] However, in ZrGeSe, the total resistivity magnitude change is quite small, which may effectively explained by Kohler's rule. We notice that in semimetal system like WTe$_2$,[42] or PtSn$_4$ & PdSn$_4$,[43] Kohler's rule is successful to attribute the field induced low temperature $RT$ curves upturn upon cooling. Since Kohler's rule is both powerful and simplified, we employ Kohler's plot on temperature dependent magnetoresistance (which will be mainly discussed in next section) shown in **Fig. S6 Supporting Information**. We found that, in a certain degree, Kohler's rule scaling works well at 3 – 300 K region, indicating that the carrier's scattering mechanism may be the same at different temperatures. In this case, even no field-induced energy gap opens, $RT$ curves still present local minimums at high magnetic field region. We would like to leave this discussion as an open question in ZrGeSe system, a further low temperature & high magnetic field STM study may solve this puzzle.

2）**Magnetoresistance and SdH Quantum Oscillations**

The transverse magnetoresistance ($MR$) is the change of resistance with the magnetic field applied perpendicular to the electric current, which is defined by $MR = \frac{\rho(B)-\rho(0)}{\rho(0)}$, where $\rho(H)$ is the resistivity in an applied magnetic field $H$. Since $MR$ measurements could reflect the Shubnikov-de Haas (SdH) oscillations and help to gain insight into the size and shape of the Fermi surface of semimetals. We thus conduct the *MR* measurements at several fixed temperatures and plotted *MR* as a function of magnetic field in **Fig. 3a**. At low temperatures ($T \leq 30$ K), large, non-saturating *MR* are obtained, reaching 1860% at 3 K and 14 T. For $T \geq 100$ K the variation of *MR* with the magnetic field is quite similar to that observed in traditional metal materials. *MR* decreases to 11% at $T=300$ K and $B=14$ T. At $T=3$ K the *MR* curve follows a quadratic field dependent behavior at low-field region, while almost linear at high-field region. Interestingly, one can find trace of resistivity oscillations in the low-temperature region, e.g., 3 – 30 K, which is known as SdH oscillations. To extract the oscillatory component $\Delta\rho$, a smooth background is subtracted from the *MR* curves. As shown in **Fig. 3b**, the oscillations are still clear at 30 K. One can find that the oscillation patterns which show obviously multi-frequency behaviors are of the same frequencies and phase at different temperatures. We employed fast Fourier transform (FFT) to analyze the oscillation frequencies, which are shown in **Fig. 3c**. The SdH oscillations are composed of one lower frequency $F_\alpha$ (=44.3 T) and a higher frequency $F_\beta$ (= 228.3 T) for ***B*** // *c* axis. The coexistence of lower and higher frequencies has also been observed in ZrSiS and HfSiS,[23-27] which suggests that two pockets with obvious different size can be detected from the *c* direction in ZrSiS-type nodal-line semimetals. In Hu's work,[31] one more pocket near 400 T also contributes to dHvA oscillations, which is absence in our experiment. The missing pocket in the present SdH oscillations is probably due to the lower (than Hu's work) applied magnetic field, which is failure to detect the large pocket.

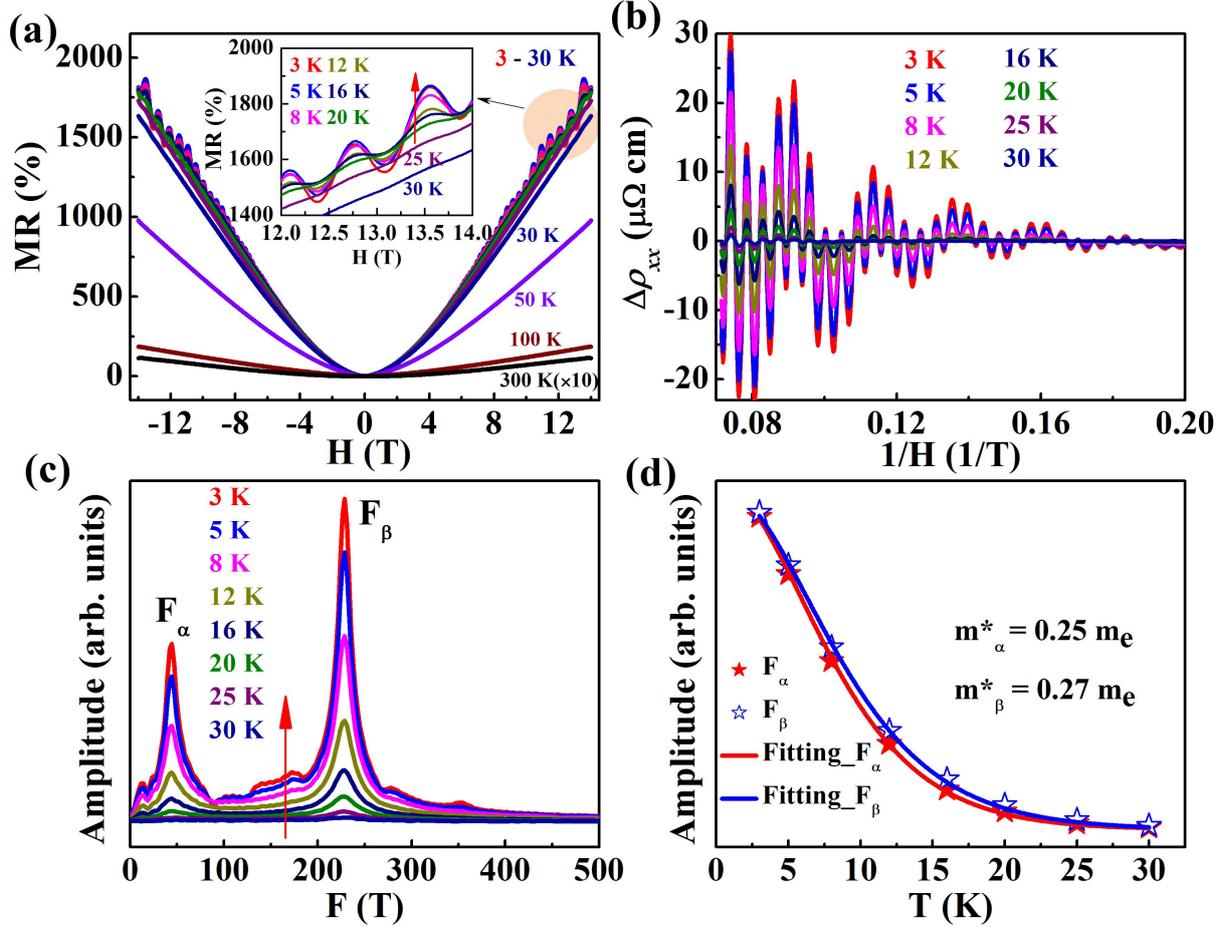

**Fig. 3 Magnetoresistance and SdH oscillations at different fixed temperatures.** (a) Magnetoresistance plotted as a function of magnetic field at different fixed temperatures for the ZrGeSe crystal. Note that the *MR* curve at 300 K is enlarged by 10 times to present the parabolic behaviors. Insets: zoom-in plot of the oscillation patterns from 3 to 30 K. (b) The oscillation patterns plotted as a function of $1/T$ by extracting from the MR data in panel (a). (c) FFT spectra with two oscillation frequencies for the SdH oscillations. (d) FFT amplitudes fitting using the L-K formula, to obtain the effective mass of the carriers.

The SdH oscillations for a semimetal can be described by the Lifshitz-Kosevich (L-K) formula,[44] with the Berry phase being taken into account for a topological system:[45]

$$\frac{\Delta\rho}{\rho(0)} = \frac{5}{2}(\frac{B}{2F})^{\frac{1}{2}}R_T R_D R_S \cos\left[2\pi(\frac{F}{B}+\gamma-\delta)\right]$$

where $R_T = \alpha T\mu/B\sinh(\alpha T\mu/B)$, $R_D = \exp(-\alpha T_D\mu/B)$, and $R_S = \cos(\alpha g\mu/2)$. Here, $\mu = m^*/m_0$ is the ratio of effective cyclotron mass $m^*$ to free electron mass $m_0$; $T_D$ is the Dingle temperature, and $\alpha = (2\pi^2 k_B m_0)/\hbar e$. The oscillations of $\Delta\rho$ is described by the sine term with a phase factor $\gamma - \delta$, in which $\gamma = 1/2 - \Phi_B/2$, and $\Phi_B$ is the Berry phase. From the L-K formula, the effective mass of carrier contributing to the SdH effect can be obtained through the fitting of the temperature dependence of oscillation amplitude to the thermal damping factor $R_T$. In the case of multi-frequency oscillations, the oscillation amplitude for each frequency can be represented by the amplitude of FFT peak. The parameter $1/B$ in $R_T$ should be the average inverse field $1/B$, defined as $1/B = (1/B_{max} + 1/B_{min})/2$. As shown in **Fig. 3d**, for both the low and high frequencies the obtained effective mass are $0.25m_0$ and $0.27m_0$, respectively, which are similar to the results obtained from the dHvA oscillations.[26,31] We also measured the longitude magnetoresistance with the direction of the magnetic field parallel to the current direction. The magnetoresistance is positive and one order of magnitude smaller than the transverse one, and it still shows clear signs of Shubnikov-de Haas oscillations at low temperatures (**Fig. S3, Supporting Information**).

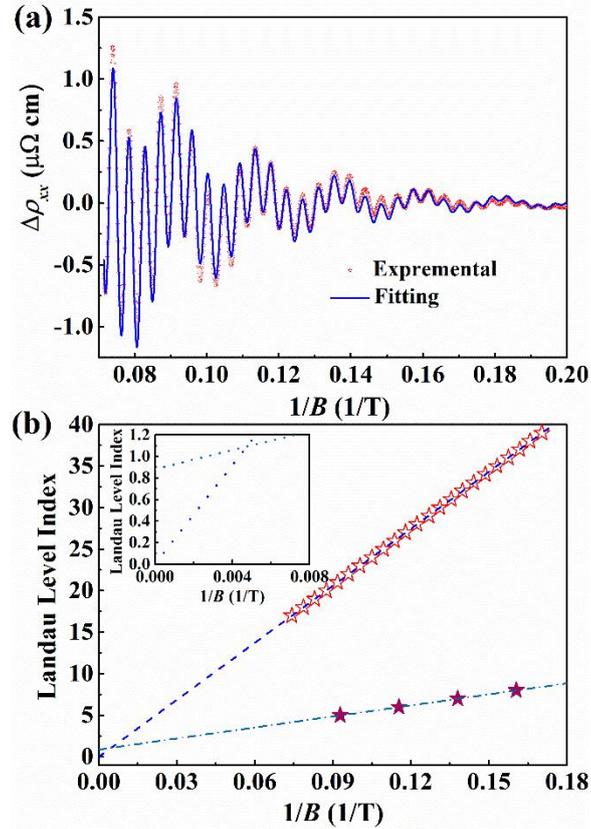

**Fig. 4 Analyses of the SdH oscillations at 3 K.** (a) Fitting of the oscillation patterns at 3 K using the L-K formula. (b) The Landau fan diagram of both oscillation patterns. Inset: zoom-in plot of linear fitting lines.

Since the multi-frequency nature of the oscillations, we fitted the oscillation patterns employing the multi-frequency L-K formula (**Fig. 4a**) and obtained the phase shift factor $\gamma - \delta$ related to the two oscillation frequencies are 0.93 and 1.08, respectively. Thus, the Berry phase would be $(0.86 \pm 0.25)\pi$ and $(1.14 \pm 0.25)\pi$, respectively. According to the Onsager-Lifshitz equation, $F = (\varphi_0/2\pi^2)A_F$, where $A_F$ is the external area of the cross section of the Fermi surface perpendicular to the magnetic field, and $\varphi_0$ is the magnetic flux quantum. The cross sections related to the 44.3 and 228.3 T pockets are $4.2\times10^{-3}$ and $21.8\times10^{-3}$ Å$^{-2}$, respectively. To determine the approximate value of the carrier density for the Fermi pockets, we used its relation with the oscillation frequency, $\Delta\left(\frac{1}{B}\right) = \frac{2e}{\hbar}(\frac{g_s g_v}{6\pi^2 n})^{2/3}$, where $g_s$ and $g_v$ are spin and valley degeneracies, respectively. From the magnetic-field-induced damping of oscillation amplitude (**Fig. 4a**), $\Delta\rho \propto \exp\left(-\frac{2\pi^2 k_B m^* T_D}{\hbar eB}\right)$, the Dingle temperature $T_D$ are 3.6 and 7.9 K which are related to the large and small Fermi pockets, respectively. To obtain a quantitative estimate of the carrier mobility, we calculated the quantum mobility, $\mu_q = e/2\pi k_B m^* T_D$. The aforementioned parameters are shown in **Table 1**.

Table 1. Parameters extracted from SdH oscillations for two Fermi pockets

| $F$ (T) | $A_F$ ($10^{-3}$ Å$^{-2}$) | $m^*$ ($m_0$) | $n_q$ ($10^{18}$ cm$^3$) | $\mu_q$ ($10^3$ cm$^2$/Vs) | $\Phi$ ($\pi$) |
|---|---|---|---|---|---|
| 44.3 | 4.2 | 0.25 | 1.6 | 2.3 | $0.86 \pm 0.25$ |
| 228.3 | 21.3 | 0.27 | 19 | 1.1 | $1.14 \pm 0.25$ |

$A_F$, and $\Phi$ are the cross section area in Fermi surface, and Berry phase of the Fermi pockets, respectively.

Another approach to obtain the Berry phase is through the Landau fan diagram which is shown in **Fig. 4b**. Since $\rho_{xx} > \rho_{xy}$, we assign the maxima of the SdH oscillations as integers (*n*) Landau level to linearly fit the *n* versus $1/H$ curve. As one can see in the inset of Fig. 4(b), the *y*-axis-intercept values of 0.85 and 0.03 are obtained for the 44.3 and 228.3 T pockets, respectively, illustrating the non-trivial Berry phase in the transport behaviors.

3）**Angle Dependent SdH Oscillations**

To further understand the geometry of Fermi surface, we performed angle-resolved SdH oscillations measurements. The *MR* versus *B* curves, as measured at different orientations of magnetic fields, are shown in **Fig. 5a**, where the magnetic field is always perpendicular to the current or the *b* axis of the crystal in order to ignore the influences of the angle between the magnetic field and the current [see the schematic diagram shown in the inset of **Fig. 5(a)**]. The rotation parameter $\theta$ is the angle between the *c* axis and the direction of the magnetic field. Namely, the rotation of the direction of magnetic field is always within *ac* plane. As the direction of the magnetic field rotates toward the *ab* plane, *MR* increases gradually before 45 degree and then decrease with further increase in $\theta$. In fact, the MR values at certain magnetic field (e.g., 14 T) forms a "butterfly" shape (**Fig. S4, Supporting Information**). The fourfold-like symmetry of the angle dependent *MR* curve at a fixed magnetic field implies that the Fermi surface of the ZrGeSe crystal is three dimensional[46]. To clearly demonstrate the oscillations in the angle dependent *MR* curves, we multiply the higher-than-50-degree data with different constants to demonstrate the increasing tendency. It can be seen that the oscillation patterns change near 45 degree. By subtracting the smooth background, we extract the oscillation patterns for all the rotation angles [**Fig. 5(b)**], which implies that the Fermi surface of ZrGeSe is formed by several components from different directions. The FFT spectra related to the oscillations are shown in **Fig 5c**. The $F_\alpha$ frequency slightly changes with the rotating angle, from 44.3 T at 0 degree to 21 T at 80 degree, while the $F_\beta$ frequency shows a 2-D like behavior, which increases roughly in the $1/\cos\theta$ form and vanishes at about 60 degree, as shown by the fitting line in **Fig 5d**. New frequencies ($F_\mu$ and $F_\nu$) can be found during the rotating process, which are slightly angle dependent. We roughly fit $F_\mu$ with cosine relationship in the 30–80 degree region. Although the two frequencies $F_\mu$ and $F_\nu$ seem like Zeeman splitting peaks, which was previously found in ZrSiS[39] under low magnetic fields, e.g., 5 T, we should be more careful on the absence of Zeeman splitting in dHvA results.[31] Therefore, we employ first principles calculations to theoretically analyze the Fermi surface of ZrGeSe.

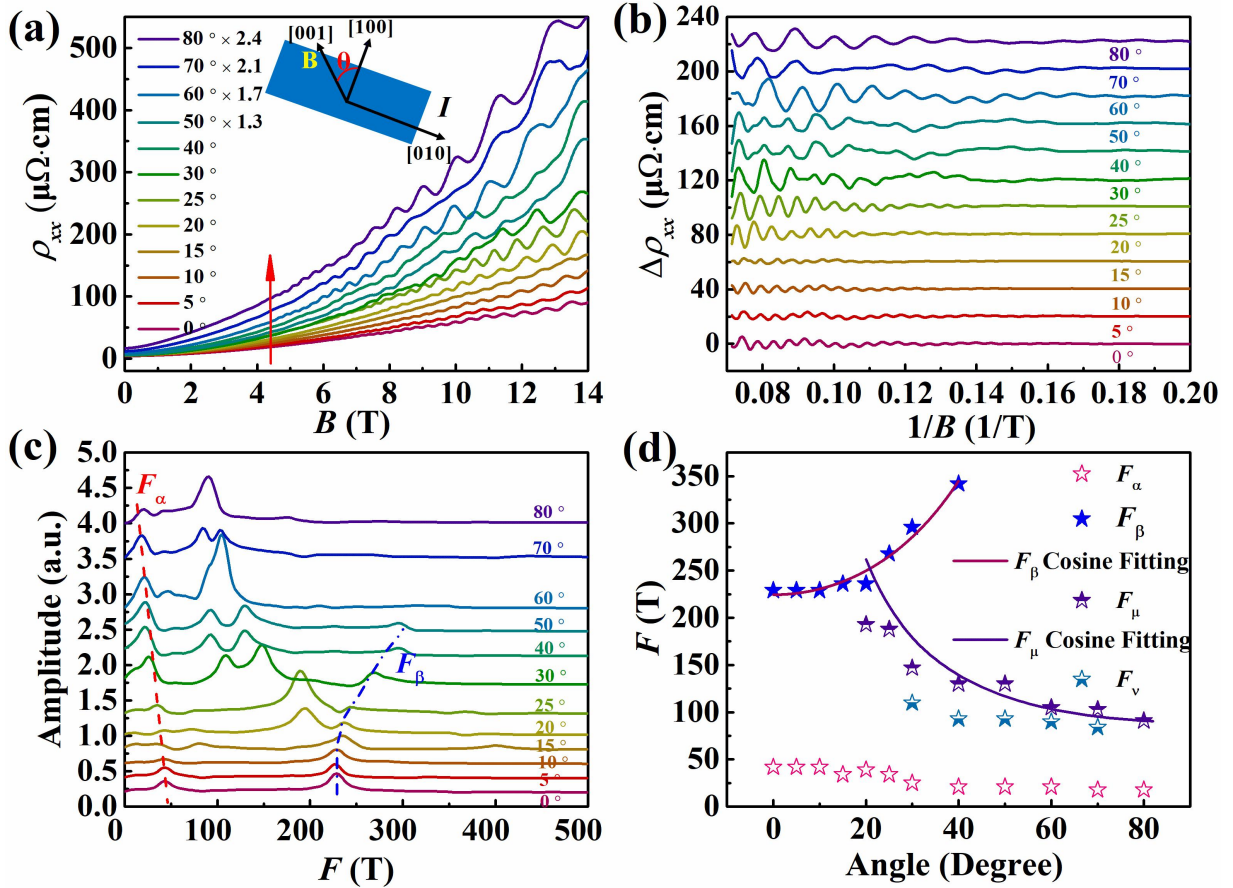

**Fig. 5 Angle dependent SdH oscillations.** (a) Resistivity plotted as a function of the magnetic field for various angles at *T*=3 K. In order to present all of the oscillation patterns without overlap, we multiply the 50–80 degree data with different constants. (b) Oscillation patterns at different angles plotted with 1/*T* by subtracting the smooth background of MR curves in panel (a). (c) The FFT patterns of the oscillation parts extracted from the MR curves. (d) The oscillation frequencies evolve with the rotation angle.

4）**DFT Calculation**

To further understand the electronic properties of the ZrGeSe crystal, first principles calculations are performed based on the density functional theory (DFT). The optimized lattice constants from the DFT calculations are $a=b=3.6891$ Å and $c=8.264$ Å. These values are quite close to the experimental ones ($a=b=3.69$ Å, $c=8.24$ Å at 300 K). According to the density of states (DOS), the bands near the Fermi level are contributed by highly hybridized

Zr's 4*d* orbitals, Ge's 4*p* orbitals and Se's 4*p* orbitals, as shown in **Fig. 6a**. The band structure displayed in **Fig. 6b** confirms the Dirac-like feature at the Fermi level. We also plot the calculated 3D Fermi surface in **Fig. 6c** which shows similar morphology with other ZrSiS-type topological nodal-line semimetals. However, an obvious difference between ZrGeSe and ZrSiS is the electron pocket at the center of the Brillouin zone $\Gamma$, which is formed by normal bands rather than Dirac bands. Upon application of magnetic fields along the *c* axis only two oscillations can be detected, which correspond to the two Dirac bands at *Z-A* and *Z-R* direction. The large pocket at *Γ*, which is, however, not even traceable, might due to the low carrier's mobility and its higher quantized field. The pocket at *Γ* brings more electron carriers in ZrGeSe, resulting ZrGeSe's electron dominant Hall behaviors shown in **Fig. S5, Supporting Information** (ZrSiS is hole dominant[39]). The obtained carriers' mobility is about 1000 cm$^2$V$^{-1}$s$^{-1}$ at low temperature from Hall measurements, which is of the same magnitude, however slightly smaller than, quantum mobility. It is reliable that the carrier's mobility would be significantly smaller than those of Dirac bands. Namely, the Dirac bands still dominant magnetotransport behaviors, resulting in the large unsaturated MR effect as well as SdH oscillations in both *MR* and Hall effects measurements.

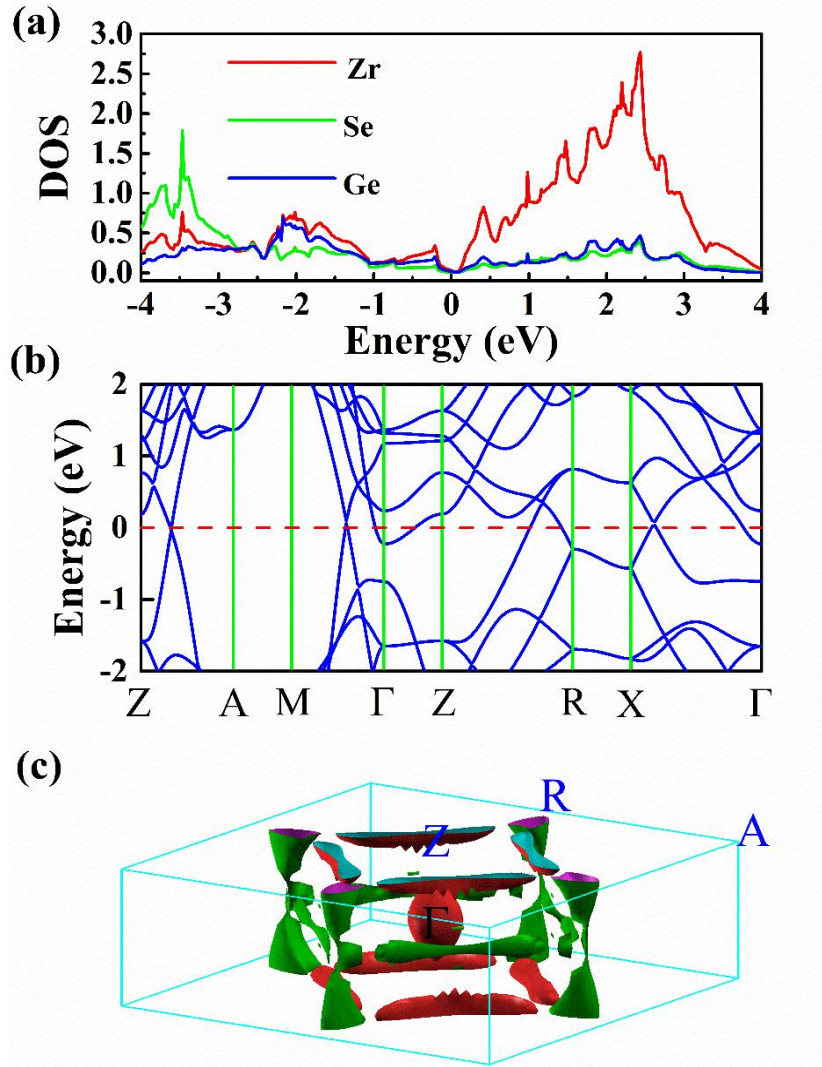

**Fig. 6. Calculated band structure and Fermi surface of ZrGeSe.** (a) Density of states. (b) Band structure. (c) 3D Fermi surfaces of ZrGeSe in the reciprocal space.

In **Fig. 6c**, the famous nodal ring of ZrSiS-type nodal-line semimetal can be found in the ZrGeSe near *Z-A-R* plane, which is more or less 2D like, and contributes to the quasi 2D like behavior in angle dependent SdH analyses. Moreover, one can find another nodal ring at the middle plane of the Brillouin zone, which is slightly different in size. This nodal ring may contribute to the multi-frequency oscillation patterns in the angle dependent experiments.

## 4. CONCLUSIONS

In summary, high quality ZrGeSe single crystals were synthesized via the chemical vapor transport. Electronic transport measurements show that ZrGeSe single crystal exhibits metallic conductivity down to 3 K and resistivity plateaus below *T*=10 K for *B*≥3 T, whose magnitude is strongly enhanced by the application of magnetic fields with its direction perpendicular to the electric current direction. Upon sweeping the magnetic field at *T*=3 K, the ZrGeSe crystals show large positive magnetoresistance (1.86 × 10$^3$%) with quantum oscillations in the high-field region. The analyses of SdH quantum oscillations data reveal properties consistent with theoretically predicted topological semimetal state. Combining angle dependent SdH oscillations and DFT calculations, we deduce that the Fermi surface of the ZrGeSe is constructed by normal bands and Dirac bands. The latter plays the dominant role in determining the magnetotransport behaviors. All these findings demonstrate that ZrGeSe is one of good nodal-line semimetals for further theoretical and experimental investigation.

## ■ ASSOCIATED CONTENT

### Supporting Information

The Supporting Information is available free of charge on the ACS Publications website at DOI: xx.xxxx/acsami.xxxxxxx. X-ray energy dispersive spectroscopy (EDS) elemental mapping of Zr, Ge, and Se, temperature dependence of the resistivity in zero magnetic field together with the fitting of the resistivity using equation $\rho(T) = \rho_0 + AT^n$, longitudinal magnetoresistance as a function of the magnetic field at different fixed temperatures, polar plots of angular dependent normalized resistivity at $T=3$ K and in magnetic fields of $B=4$, 9, and 14 T and the Hall resistance as a function of the magnetic field at different fixed temperatures for ZrGeSe single crystal. Kohler's plot of magnetoresistance at different temperatures.

## ■ AUTHOR INFORMATION

### Corresponding Authors
*E-mail: zrk@ustc.edu (R. K. Zheng)
*E-mail: sdong@seu.edu.cn (S. Dong)
*E-mail: wz929@uowmail.edu.au (W. Zhao)

### Author Contributions
$^\nabla$ L. Guo and T. W. Chen contributed equally to this work. The manuscript was written through contributions of all authors. All authors have given approval to the final version of the manuscript.

### Notes
The authors declare no competing financial interest.

## ■ ACKNOWLEDGMENTS

This work was supported by the National Natural Science Foundation of China (Grant Nos. 51572278, 51790491, 51872278, 11674055) and the National Key Research and Development Plan (Grant Nos. 2016YFA0300103 and 2015CB921201). Support from Jiangxi Key Laboratory for Two-Dimensional Materials is also acknowledged. W.Z. & L.C. acknowledge the scholarship supporting from UOW.